# Non-invasive evaluation of aortic stiffness dependence with aortic blood pressure and internal radius by Shear Wave Elastography and Ultrafast Imaging.

*C. Papadacci[1], T. Mirault[2,3], B. Dizier[3], M. Tanter[1], E. Messas[2,3,4], M. Pernot[1]*

*(1) Physics for Medicine Paris, ESPCI, CNRS, Inserm, PSL University, Paris, France*

*(2) AP-HP – Hôpital Européen Georges Pompidou, Paris, France*

*(3) Université Paris 5, Inserm U765, Paris, France*

*(4) Université Paris 5, Inserm U970, Paris, France*

# Abstract

Elastic properties of arteries have long been recognized as playing a major role in the cardiovascular system. However, non-invasive in vivo assessment of local arterial stiffness remains challenging and imprecise as current techniques rely on indirect estimates such as wall deformation or pulse wave velocity. Recently, Shear Wave Elastography (SWE) has been proposed to non-invasively assess the intrinsic arterial stiffness. In this study, we applied SWE in the abdominal aortas of rats while increasing blood pressure (BP) to investigate the dependence of shear wave speed with invasive arterial pressure and non-invasive arterial diameter measurements. A 15MHz linear array connected to an ultrafast ultrasonic scanner, set non-invasively, on the abdominal aorta of anesthetized rats (N=5) was used. The SWE acquisition followed by an ultrafast (UF) acquisition was repeated at different moment of the cardiac cycle to assess shear wave speed and arterial diameter variations respectively. Invasive arterial BP catheter placed in the carotid, allowed the accurate measurement of pressure responses to increasing does of phenylephrine infused via a venous catheter. The SWE acquisition coupled to the UF acquisition was repeated for different range of pressure. For normal range of BP, the shear wave speed was found to follow the aortic BP variation during a cardiac cycle. A minimum of (5.06±0.82) m/s during diastole and a maximum of (5.97±0.90) m/s during systole was measured. After injection of phenylephrine, a strong increase of shear wave speed (13.85±5.51) m/s was observed for a peak systolic arterial pressure of (190±10) mmHg. A non-linear relationship between shear wave speed and arterial BP was found. A complete non-invasive method was proposed to characterize the artery with shear wave speed combined with arterial diameter variations. Finally, the results were validated against two parameters the incremental elastic modulus and the pressure elastic modulus derived from BP and arterial diameter variations.

# Introduction

Arterial Stiffness is of major interest in the understanding of the cardiovascular function [1]–[3] and is widely recognized as a powerful prognostic factor of cardiovascular events. Many techniques have been developed to measure arterial stiffness non-invasively. However, arterial stiffness indexes are still not used routinely in clinical practice, in part because available techniques are not yet suitable for widespread clinical practice. Indeed, all these techniques rely on indirect estimates of the arterial stiffness through measurements of arterial pressure, distensibility or a combination, which at the end makes the arterial stiffness less accurate and more complex to estimate. Therefore, there is still a need for an easy to use, reliable and noninvasive method to quantify arterial stiffness.

The pulse wave velocity (PWV) is the most widely available technique [4]. PWV is related to the elasticity of the arterial wall because the propagation of the arterial pulse wave is driven by the ability of the arterial wall to deform when the pressure increases. Applanation tonometry [5] and oscillometric method have been used to detect the time delay of pressure waveforms at two arterial sites (generally at the carotid and femoral arteries). The PWV is then retrieved by dividing the time delay to the distance between the two sites [4]. This method is non-invasive, provides an average velocity over a long arterial distance, but suffers from inaccuracies due to the challenging measurement of the vascular length especially in the potential arterial tortuosity in older patients. Recently, real-time ultrasound imaging of the pulse wave propagation with a reduced number of transmitted beams has been proposed by Kanai et al [6] and later, extensively studied by other groups to estimate the local PWV [7]–[12]. However, although the PWV has been used successfully for decades as an arterial stiffness index, it remains an indirect measurement of the arterial stiffness which depends strongly on the size of arteries (diameter and wall thickness). For example, the PWV of arteries of different size should not be used to compare their arterial stiffness.

Local arterial stiffness can also be assessed by measuring the distensibility of the arterial wall using real-time ultrasound imaging. The technique known as echotracking has been developed to measure finely the change of diameter of the artery during the cardiac cycle [13]. The arterial wall distensibility can then be computed by using the diameter and the blood pressure. As a matter of fact, this approach requires the knowledge of the intraluminal arterial pressure which cannot be measured directly. Pressure of superficial arteries is usually estimated using aplanation tonometry, but for deep arteries such as aorta, a transfer function must be applied to estimate the blood pressure. These transfer functions are established on average populations, however, there is a large scatter in measured transfer functions so that central pressure may differ considerably from the direct measurement [14],[15].

In another hand, Shear wave elastography (SWE) has been proposed to measure directly, locally, non-invasively and in real time the elasticity of soft tissues [16]. Shear waves are generated by the acoustic radiation force of a focused ultrasonic beam emitted by an ultrasonic probe. By imaging the propagation of the shear wave at very high frame rate using ultrafast ultrasound imaging, the shear wave speed can be quantified locally, and the local shear modulus can be derived. This technique has been successfully implemented to many organs such as the breast [16], the liver [17], the cornea [18], the tendon [19] and the heart [20]–[22]. We previously proposed to apply this technique for the estimation of the arterial stiffness in Couade et al. [23]. We showed that SWE has the potential to measure quantitatively, locally and non-invasively the stiffness of the arterial wall and its variation during the cardiac cycles. However, the technique was not yet compared against other gold standard measurements and no index was proposed to be used in the clinic.

In this study, we propose to measure arterial stiffness using SWE over a large range of arterial pressure (40 mmHg-190 mmHg) and compare it with gold standard measurements. The aortic stiffness of rats (N=5) was measured non-invasively by SWE as well as the artery diameter by ultrafast imaging. Simultaneously, the blood pressure was assessed using invasive pressure catheter. Arterial blood pressure was varied using phenylephrine and the incremental Young's modulus was obtained using both invasive and non-invasive methods. Finally, to validate the qualitative mechanical behavior of the artery, shear wave speed versus diameter curves obtained non-invasively were compared to the incremental elastic modulus and the pressure elastic modulus obtained with invasive measurements.

# Material and methods

*1) Stiffness evaluation in artery with SWE*

SWE has been extensively studied in the past years. It relies on the creation of an acoustic ultrasonic burst emitted from a focused beam of an ultrasonic probe which produces pressure radiation acoustic force. The pressure radiation force locally pushes the tissue and its relaxation produces a shear wave propagating at a speed $c$ in the medium. In soft tissues, it has been demonstrated that the elastic modulus (Young's modulus) $E$ which represents stiffness was only a function of shear modulus µ in soft isotropic tissues as:

$$E \approx 3\mu \qquad (1)$$

and shear modulus µ related to shear wave speed c and density $\rho$ of the soft tissue as:

$$\mu \approx \rho\, c^2 \qquad (2)$$

The shear wave speed can then be assessed by tracking the shear wave propagation with ultrafast plane wave imaging [16]. While arterial wall is a soft tissue like the liver or the breast, it is rather more complicated to assess its stiffness.

Firstly, because the elastic modulus is a function of stress and strain. In static soft tissues such as the breast or the liver, the stress and strain remains the same which implies a unique measurement under one strain-stress condition. However, in artery the stress applied on the arterial wall varies naturally and dynamically in a cardiac cycle due to the internal blood pressure variation. The strain follows the stress variation by modifying the arterial wall thickness. It is known that the relationship between stress and strain in biological tissue is a non-linear relationship [24]. Therefore, an assessment of elasticity at different blood pressure values (or different stress values) is necessary to assess the mechanical behavior of the artery and to characterize it under different stress-strain conditions.

Secondly, because the arterial wall is an anisotropic soft tissue, the elastic modulus is a tensor with different components. The configuration shown in figure 1 (a),(b),(c), was the configuration of the shear wave elastography acquisition in this study. The stiffness along the longitudinal direction $\hat{u}_z$ (figure 1 (a)) could be different from the stiffness assessed in the circumferential direction $\hat{u}_\theta$.

Finally, because the arterial wall is small compared to the shear wavelength, the shear waves are guided in the arterial wall. The guided wave propagation model to retrieve the quantitative elastic modulus is complex and depends on the shear wave central frequency, the cylindrical geometry of the artery and the ratio between the wall thickness and the radius of the artery. In this study, shear wave speeds were used as a stiffness index with no attempts to derive the shear modulus.

*2) Experimental Setup*

A high frequency ultrasound linear probe (15 MHz, 128 elements, 0.125 mm pitch, Vermon, France) connected to an ultrasonic scanner (Aixplorer, SuperSonic Imagine, Fance) was positioned non-invasively, on the abdominal aorta of anesthetized rats (N=5). 2 pressure catheters were set invasively in the left common carotid artery and in the left femoral artery. Blood pressure at the two sites were recorded in real time (figure 2, green and black line). ECG was recorded via 4 needle electrodes implanted in the legs (figure 2, blue line). The echograph was trigged on the R-wave from ECG. Trig out of the echograph was also recorded to verify that SWE acquisitions and ultrafast acquisitions occurred at the desired time (figure 2, red line).

Pressure from the two sites were delayed and averaged to estimate the blood pressure at the location of the SWE acquisition. The blood pressure was varied from about 40 mmHg to 190 mmHg using a vasoconstrictor (phenylephrine) infused via a venous catheter.

*3) Data acquisition*
a)     *SWE sequence*

Longitudinal shear wave speed was assessed during SWE sequence. One pattern of the SWE acquisition consisted in the emission of one pushing beam focused at the superior arterial wall (between 4 mm and 6 mm depth depending on the animal) during 300 µs, at the probe center with an F over D (focal distance over aperture) of 1.5. 80 tilted plane waves (with angles of [-2° 2°]) were transmitted at the frame rate of 31,000 images/s in order to coherently sum 40 compounded plane waves with a frame rate of 15,500 images/s according to [25]. This pattern was repeated 10 times every 15 ms to cover an entire cardiac cycle. This SWE sequence was trigged on the R-wave of the ECG (i.e. figure 3 a).

b)     *Ultrafast plane wave acquisition*

Non-invasive evaluation of diameter variation was performed using ultrafast plane wave acquisition data. After SWE acquisitions, independent ultrafast plane wave imaging acquisitions were performed for identical value of blood pressure (i.e. figure 3 b). The acquisitions were also triggered on the R-wave of the ECG. The ultrafast imaging acquisition consisted in the emission of 4500 tilted plane waves (with angles of [-2° 0 2°]) at a frame rate of 27,000 images/s in order to achieve 1500 compounded plane waves at a frame rate of 9,000 images/s. The ultrafast acquisition lasted 166 ms which corresponded to one rat cardiac cycle.

*4) Postprocessing*
a)     *SWE data*

IQ signals from the 80 tilted plane waves were stored in memory. Coherent compounding was performed to get 40 compounded images. B-mode images were retrieved from absolute value of compounded IQ signals (figure 4 a). Tissue velocities (figure 4 b) were obtained using a per pixel frame to frame 1D cross-correlation on demodulated compounded IQ images with an axial kernel size of 3 pixels (150 µm) to obtain images of tissue frame-to-frame axial displacements. The superior arterial wall was selected manually (figure 4 a, blue line). Tissue velocity was averaged from -1 pixel to 1 pixel in depth around the selected line and displayed as a function of time (figure 4 b). Shear wave speed was assessed by fitting the maxima of tissue velocity of the shear wave front with a linear fit (figure 4 b, white line). The slope of the fit gave access to the longitudinal speed of the shear wave.

b)     UF data

IQ signals from the 4500 tilted plane waves were stored in memory. Coherent compounding was performed on the three angles to get 1500 compounded images.

Similarly to SWE post processing, tissue velocities (figure 4 b) were obtained using a per pixel frame to frame 1D cross-correlation on demodulated compounded IQ images with an axial kernel size of 3 pixels (150 µm) to obtain images of tissue frame-to-frame axial displacements. Internal diameter $d_0$ and wall thickness $h_0$ of the artery at a reference time $t_0$ were measured on the focused B-mode images of the SSI echograph (i.e. figure 5 a). Arterial wall radial displacement of the anterior wall $\delta d_1$ and posterior wall $\delta d_2$ were obtained with ultrafast sequence by integrating tissue velocities over time of the anterior wall and posterior wall respectively (i.e. figure 5 b). Internal diameter variation was therefore computed as:

$$\delta d = \delta d_1 - \delta d_2 \tag{3}$$

Internal diameter $d$ and internal radius $a$ as a function of time was calculated as:

$$d = (d_0 + \delta d) \, ; \, a = \frac{d}{2} \tag{4}$$

Under incompressibility assumption, wall thickness $h$ as a function of time was derived by resolving the equation:

$$(\frac{d}{2} + h)^2 - \frac{d^2}{4} = C_0 \tag{5}$$

With the constant $C_0$:

$$C_0 = (\frac{d_0}{2} + h_0)^2 - \frac{d_0^2}{4} \tag{6}$$

Finally, the external radius $b$ was found as:

$$b = a + h \tag{7}$$

5)  *Comparison with validated techniques*

We compared the longitudinal elastic modulus values estimated by SWE to the circumferential elastic modulus assessed by two validated methods: the pressure elastic modulus [26] and the incremental elastic modulus [27]–[29] derived from pressure and distensibility (relative cross-sectional diameter change for a given pressure) measurements. These two methods are based on stress and strain measurements.

a)     The pressure elastic modulus

Pressure Young's modulus $E_{P\theta}$ was defined by [26]:

$$E_{P\theta} = P \frac{a^2}{h \, \delta a} \tag{8}$$

It was estimated by measuring invasively the blood pressure $P$ with the catheters, the internal radius $a$, the arterial radius variation $\delta a$ and the wall thickness $h$ from ultrafast acquisitions data.

b)     The incremental elastic modulus

This method also assumes homogeneity and isotropy of the material. The incremental elastic modulus was calculated from the pressure-radius data as:

$$E_{inc\theta} = \frac{3}{2} \frac{a^2 b}{(b^2 - a^2)} \frac{\Delta P}{\Delta b} \tag{9}$$

Where $\frac{\Delta P}{\Delta b}$ represents the slope of the pressure-external radius curve at a specific point (i.e., pressure or strain). Incremental elastic modulus at a point was determined by calculating the derivative of the pressure-external radius curve considering the data points below and above.

# Results

1) Shear wave speed over one cardiac cycle

At each SWE acquisition, shear wave propagation could be tracked in the superior arterial wall of the aorta with high temporal resolution. Figure 6 displays snapshots of the shear wave propagation superimposed on the B-mode image at different time. Shear waves appeared to propagate faster in the arterial wall than in the surrounding tissues.

SWE acquisition had the sensitivity to measure the variation of shear wave speed associated to the natural variation of arterial blood pressure during one cardiac cycle. Figure 7 a b, displays an example of shear wave speed estimations at ten different moment of one cardiac cycle of a rat as a function of time. In this example, the blood pressure varied between 92 mmHg in diastole to 130 mmHg in systole (i.e. figure 7 d). The shear wave speeds were found to follow the blood pressure variations from 5.1 m/s in diastole to 10.2 m/s in systole. The internal radius variation was also found to follow the blood pressure variation from 811 µm to 863 µm (figure 7 c).

2) Stiffness as a function of pressure

Arterial blood pressure was gradually varied from 40 mmHg (minimum in diastole) to 190 mmHg (maximum in systole) for the N=5 rats. For each set of pressure, shear wave speeds were assessed simultaneously. Figure 8 displays the results of shear wave speed as a function of invasive measure of blood pressure for the 5 rats. The shear wave speeds were averaged over the 5 rats over 10 mmHg step from 40 mmHg to 190 mmHg, error bars were computed as the standard deviation. Arterial stiffness was found to increase with pressure and presents two different behavior depending on the pressure range. First, a small increase of shear wave speed (slope $1.9\ 10^{-2} m/s.mmHg^{-1}$) from (5.06±0.82) m/s to (5.97±0.90) m/s was found for pressure variation from (40±5) mmHg to (95±5) mmHg respectively. For higher pressure range ((105±5) mmHg-(190±5) mmHg), a large increase (slope $8.8\ 10^{-2} m/s.mmHg^{-1}$) of shear speeds from (6.90±1.45) m/s to (13.85±5.51) m/s was shown. These two parts evidenced the non-linear behavior of the artery's elasticity under a radial stress. Two linear fits were applied to the two curve parts showing a large slope difference.

3) Comparison of shear wave speed against internal radius with validated techniques.

Shear wave speeds measured non-invasively with the SWE method were compared against internal radius. The results are shown in figure 9 for N=3 rats. Once again, the curve is divided in two parts. First, a small increase (slope of $18.2\ m/s.mm^{-1}$) of shear wave speed (5 m/s to 5.9 m/s) was found for internal radius variation from 798 µm to 835 µm. For larger radius range (838 µm-862 µm), a large increase (slope of $174\ m/s.mm^{-1}$) of shear wave speed was found (5.8 m/s to 9.9 m/s).

To compare these results, incremental elastic modulus and pressure elastic modulus estimations were performed. Measure of internal radius, external radius and invasive measurement of blood pressure were combined in order to estimate the incremental elastic moduli with respect to equation 9 and to estimate pressure elastic moduli (equation 8).

Results are shown in figure 9 for N=3 rats. The incremental and pressure elastic moduli are shown as a function of internal radius (curve blue and black respectively). Similarly, to the SWE results, two parts can be separated. The first parts of the two curves corresponded to a small increase from 166 kPa to 309 kPa for the incremental elastic moduli and from 154 kPa to 246 kPa for the pressure elastic moduli, for radii values between 798 µm and 835 µm respectively. A larger increase is then observed for internal radius variation from 798 µm to 835 µm. Interestingly, the three methods gave very close qualitative results. Although, the values of the three methods could not be compared, the mechanical behavior of the artery remained the same at small and large radius.

## Discussion

In this study, SWE was performed in the aortas of N=5 rats. Shear wave propagation was observed on the anterior wall of the rats' aorta (figure 6 a b c) at ten different moments during one cardiac cycle. Shear wave speed variation due to the physiological blood pressure change was measured (figure 7 a b). Two intravenous catheters were inserted on the two edge of the abdominal aorta to monitor and record blood pressure variations in real-time (figure 7 d). UF imaging was also performed to evaluate the internal radius variation (figure 7 c) which was found to follow shear wave speed (figure 7 b) and blood pressure (figure 7 d).

Shear wave speed in the abdominal aortic wall of N=5 rats was then measured over a large range of arterial pressure by increasing the mean blood pressure using phenylephrine injection. With the invasive pressure measurements, the relationship between shear wave speed and blood pressure was assessed. An increase of shear wave speed with the pressure was observed (figure 8). This increase is due to the fact that the blood pressure level is proportional to the stress induced on the arterial wall which is also proportional to the elastic modulus. Therefore, when the blood pressure increases, the shear wave speed increases as well.

However, we observed a non-linear relationship between shear wave speed and blood pressure level indicating of a non-linear mechanical behavior. The curve can be separated in two parts. A first part can be observed at low stress between 40 mmHg and 100 mmHg as shear wave speed increases linearly and slowly with the blood pressure level with a slope of $1.9 \; 10^{-2} m/s.mmHg^{-1}$. A second part at higher stress between 105 mmHg and 195 mmHg, still displays a linear variation but with a much larger increase of shear wave speed with blood pressure level (slope of $8.8 \; 10^{-2} m/s.mmHg^{-1}$ ). These results indicate two different mechanical behaviors of the artery which depends on the level of stress. This particular behavior could be associated with the elastin-collagen model. At low stress (BP<100 mmHg), elastin fibers are recruited to stiffen the arteries while stress is increasing [30]. At higher stress (BP>100 mmHg), when elastin fibers are completely stretched, collagen fibers are recruited for the stiffening of the arteries [31]. These two fibers have different mechanical properties explaining that their elasticity does not respond identically to an increase of stress. This model could explain the difference in slopes observed on the plot of the figure 8. In addition, the slope corresponding to the collagen fibers was found larger than the one corresponding to the elastin fibers which was in a good agreement with literature.

In the third part of the study, shear wave speed against arterial internal radius was assessed and compared to the incremental elastic modulus and the pressure elastic modulus (figure 9). As expected, shear wave speed which is related to stiffness, was found to increase with internal radius. Similarly to the previous part of the study, a non-linear behavior was found and two parts of the plot could be separated. For a radius of 798 µm to 835 µm, a small linear increase was found (slope of $18.2 \; m/s.mm^{-1}$). For a larger radius from 838 µm to 862 µm, a larger linear increase was noticed (slope of $174 \; m/s.mm^{-1}$). This non-linear mechanical behavior was also associated to the elastin-collagen model. Small radius corresponding to the recruitment of elastin fibers while at larger radius elastin fibers are completely stretched and collagen fibers are used. Interestingly, we found a good agreement with the two other techniques. Although, the values cannot be directly compared as the two invasive methods give an estimation of the elastic modulus, the same behavior can nevertheless be identified. A small increase was measured for the three methods for radius from 798 µm to 835 µm indicating of an elastin fibers stiffening. For the same value of radius (838 µm), the mechanical behavior of the artery changed with the recruitment of collagen fibers.

In this study, the elastic modulus was not derived with shear wave speed. However, although the proposed method does not evaluate quantitatively the elasticity via the elastic modulus, the shear velocity is an interesting stiffness index to investigate the mechanical behavior of the artery. It is also completely non-invasive compared to the incremental and pressure elastic modulii that need an invasive measurement of the blood pressure to assess the elasticity accurately. With SWE and UF sequences the proposed method can obtain the two slopes of the curve which could be used as an index to characterize the state of health of an artery. In Ehlers-Danlos genetic disease for instance, the artery has a collagen failure. As a consequence, the artery does not stiffen efficiently at high pressure resulting in the creation of aneurysms and the death of the patients. There is currently no imaging methods that can detect and diagnose these patients. The proposed method could be of a great interest to detect and characterize the Ehlers-Danlos patients by evaluating the slope of the second part of the shear wave speed/radius curve corresponding to the mechanical properties of the collagen.

In this study, the two measurements of shear wave speed and radius evaluation were performed separately over two successive cardiac cycles, but it may be possible to combine these two sequences into one and provide the stiffness-diameter relationship in real-time.

However, the proposed method still needs the measurement of the internal and external radius at a reference time to assess radius variations. This measurement performed on the focused B-mode image can be inaccurate for several reasons. First of all, the probe can be not perfectly set across the artery, inducing an underestimation of the diameter. An ultrasonic 3D ultrafast image of the artery could be used to overcome this issue [32]–[34]. Finally, the method could provide two index to characterize the mechanical properties in arteries of patients in the clinic.

# Conclusion

In this study, we measured shear wave speed in the arterial walls of rats' abdominal aortas with shear wave elastography within entire heart cycles. Shear wave elastography was able to follow the shear wave speed variation with the natural arterial blood pressure and diameter variations. After increasing blood pressure with phenylephrine injections, the elasticity of the artery increased as a response to higher stress. Shear wave speed was measured over a large range of pressure (from 40 mmHg to 190 mmHg), and a non-linear relationship between pressure and shear wave speed was found. We also proposed a method totally non-invasive, which relies on the estimation of shear wave speed and the measurement of arterial diameter with ultrafast plane wave imaging technique. This method enabled the estimation of the arterial diameter variation with high temporal resolution through an entire cardiac cycle. The shear wave speeds and the radius were coupled to highlight the non-linear properties of the abdominal aorta and calculate the slopes associated with elastin and collagen stiffening. The proposed technique was successfully compared to two other validated techniques which relies on invasive blood pressure measurements. The slopes derived from the proposed method could be a useful index to characterize arteries completely non-invasively in clinics without the need to use blood pressure measurements.